\documentclass[a4paper,twoside]{article}

\usepackage{epsfig}
\usepackage{subcaption}
\usepackage{calc}
\usepackage{amssymb}
\usepackage{amstext}
\usepackage{amsmath}
\usepackage{amsthm}
\usepackage{multicol}
\usepackage{pslatex}
\usepackage{apalike}
\usepackage[bottom]{footmisc}
\usepackage{listings} 
\usepackage{graphicx} 
\graphicspath{{figs/}} 
\usepackage{float} 
\usepackage{url} 
\usepackage{SCITEPRESS}     

\begin{document}

\title{Functional Component Descriptions for Electrical Circuits based on Semantic Technology Reasoning}

\author{\authorname{Johannes Bayer\sup{1}\orcidAuthor{0000-0002-0728-8735},
                    Mina Karami Zadeh\sup{1}\orcidAuthor{0000-0002-6965-5190},
                    Markus Schr\"oder\sup{1}\orcidAuthor{0000-0001-8416-0535}
                    and Andreas Dengel\sup{1}\orcidAuthor{0000-0002-6100-8255}}
\affiliation{\sup{1}Deutsches Forschungszentrum f\"ur K\"unstliche Intelligenz, Trippstadter Str. 122, Kaiserslautern, Germany}
\email{\{johannes.bayer, mina.karami\_zadeh, markus.schroeder, andreas.dengel\}@dfki.de}
}


\abstract{Circuit diagrams have been used in electrical engineering for decades to describe the wiring of devices and facilities. They depict electrical components in a symbolic and graph-based manner. While the circuit design is usually performed electronically, there are still legacy paper-based diagrams that require digitization in order to be used in CAE systems. Generally, knowledge on specific circuits may be lost between engineering projects, making it hard for domain novices to understand a given circuit design. The graph-based nature of these documents can be exploited by semantic technology-based reasoning in order to generate human-understandable descriptions of their functional principles. More precisely, each electrical component (e.g. a \texttt{diode}) of a circuit may be assigned a high-level function label which describes its purpose within the device (e.g. \texttt{flyback diode for reverse voltage protection}). In this paper, forward chaining rules are used for such a generation. The described approach is applicable for both CAE-based circuits as well as raw circuits yielded by an image understanding pipeline. The viability of the approach is demonstrated by application to an existing set of circuits.}

\onecolumn \maketitle \normalsize \setcounter{footnote}{0} \vfill

\section{\uppercase{Introduction}}
\label{sec:introduction}

Graphs are a traditional mean to describe electrical circuits \cite{rucker2012network}. Computer-aided engineering (CAE) systems are already used to capture, maintain, simulate and verify these circuits by exploiting their underlying graph structure. However, circuits also incorporate engineering knowledge. During the migration of circuits between CAE systems or during the digitization of circuits from paper sources, maintaining the plain syntactic features of the graph structure is often focused while high-level functional principles are not properly processed. Likewise, errors in the migration process often need to be traced manually.

In order to help developers (and other agents) understand circuit functionality and to automatically search for engineering concepts, additional means are required. This can also be achieved by exploiting graph structures: For example, a diode which has its anode connected to a terminal of an inductor and its cathode connected to the opposite terminal of the same inductor can be considered a flyback diode (functional description of the diode component). Likewise, a direct electrical connection between the two terminals of a voltage source can be considered a shortcut (fault description). Modeling these engineering concepts by the semantic technologies in order to reason about their presence in arbitrary circuits is the objective of the paper at hand. 

\section{\uppercase{Related Work}}
\label{sec:related}
Electrical Rule checker are already existing for CAE software (e.g. \cite{kibot}).

Efforts have already been made to model electrical power systems for the purpose of interoperability \cite{gaha2013ontology} as well as fault diagnosis \cite{501068}. In contrast, the paper at hand focuses on electronic circuits.

\cite{liu1990shifting} describe a system that integrates physical aspects and statements about circuits, hence allow to question about the low-level behavior of the system. Conversely, \cite{kitamura1998functional} describe an approach for assigning functions to components of technical systems. However, the approach is rather generic way and demonstrated by the example of a power plant.

\cite{kleer1979causal} \cite{de1984circuits} describes a comprehensive system for automatically deriving high-level insights on electronic circuits. Due to the time of their implementation, the described systems lack support of modern semantic technologies like RDF \cite{rdf}.

\cite{kunal2020gana} describe an approach for sub-circuit annotation to based on graph neural networks. While this approach is powerful, it comes with the typical limitation of ANNs like a fixed class set the requirement for a (large and usually non-public) dataset as well as a lack of perfect accuracy and explainability.

Systems for the manipulation of piping and instrumentation diagrams (which are structurally similar to circuit diagrams) have been proposed by \cite{gruner2014rule} and \cite{bayer2020graph}.

An extended version of a public dataset of hand-drawn circuit diagrams \cite{thoma2021public} has been used to evaluate the approach described in this paper.

Wikidata is an open knowledge database, in which entities of many domains as well as general-purpose  properties are available\cite{van2019wikidata}. As it contains entries for describing electric and electronic components, the RDF circuit representations described in this paper is linked against this database in order to allow for later inference of additional knowledge.

\section{\uppercase{Approach}}
\label{sec:approach}

\begin{figure}
  \centering
  \includegraphics[width=0.48\textwidth]{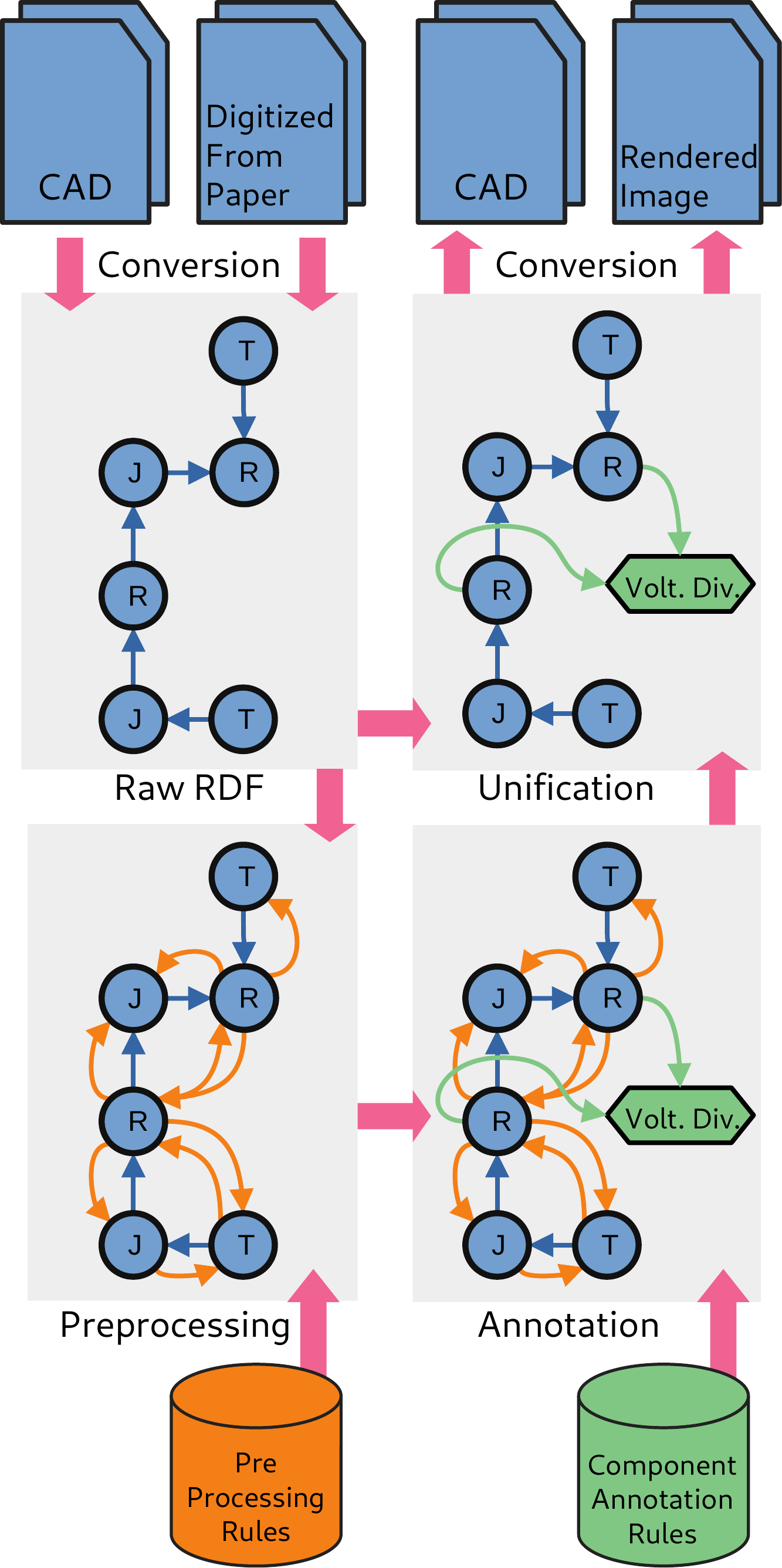}
  \caption{System Overview}
  \label{fig:overview}
\end{figure}

The complete approach consists of the following steps (see Figure~\ref{fig:overview}):

\begin{itemize}
 \item Conversion from CAE file formats or recognition results to a uniform raw RDF representation.
 \item \textbf{Preprocessing Rules} are applied in order to abstract from the circuit's optical features like junctions and to compensate structural weaknesses like directed component connections.
 \item \textbf{Component Annotation Rules} are applied to generate functional descriptions within the circuit.
 \item For avoiding redundancies, the resulting RDF representation is generated from a unification of the raw RDF and the functional descriptions.
 \item The enriched RDF is converted back to CAE or image files
\end{itemize}

Both preprocessing rules and component annotation rules are denoted in Apache Jena \cite{siemer2019exploring} forward chaining syntax.

\subsection{Ontology}

\begin{figure}
  \centering
  \includegraphics[width=0.42\textwidth]{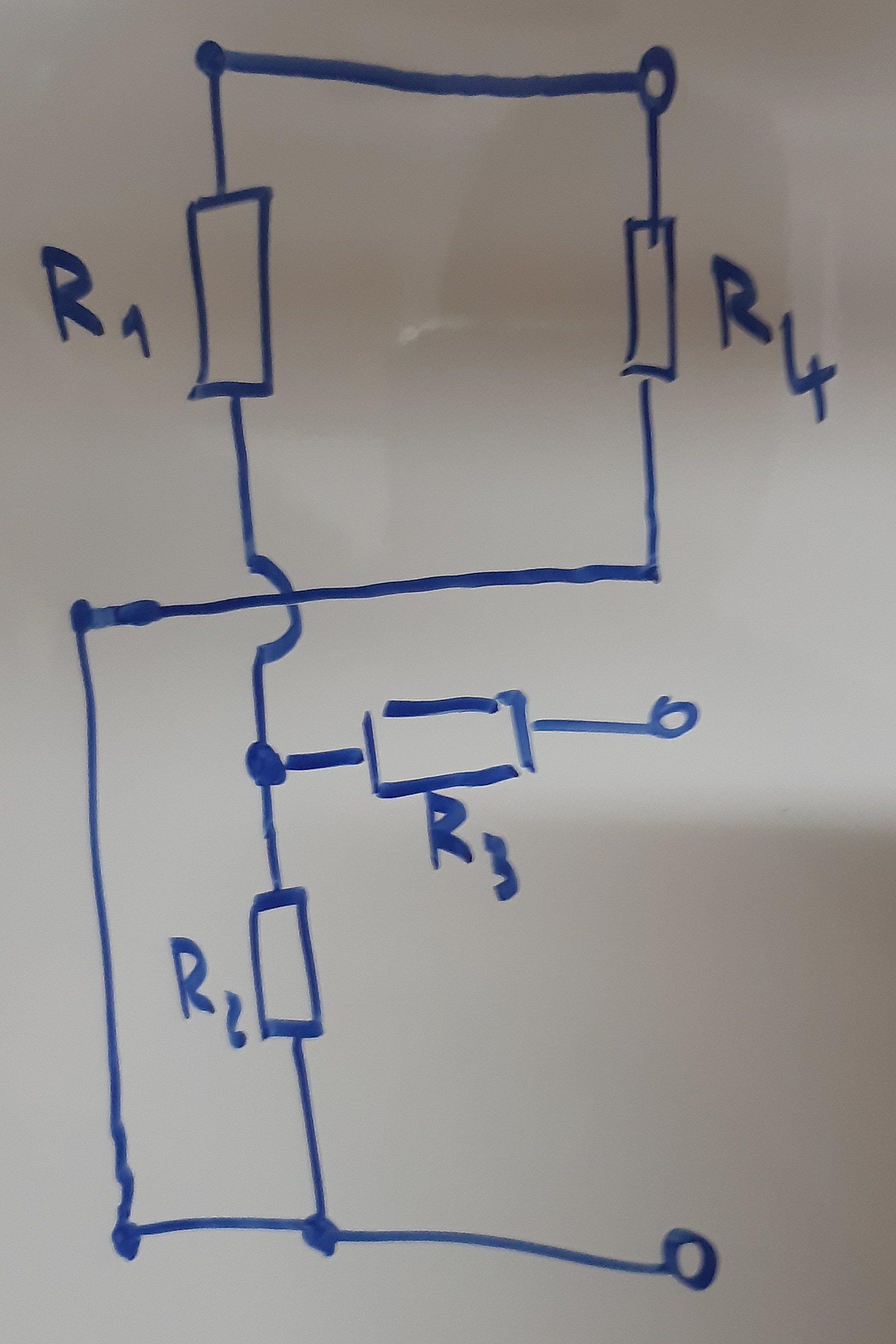}
  \caption{Sample Circuit Image.}
  \label{fig:sample_raw}
\end{figure}

\begin{figure}
  \centering
  \includegraphics[width=0.42\textwidth]{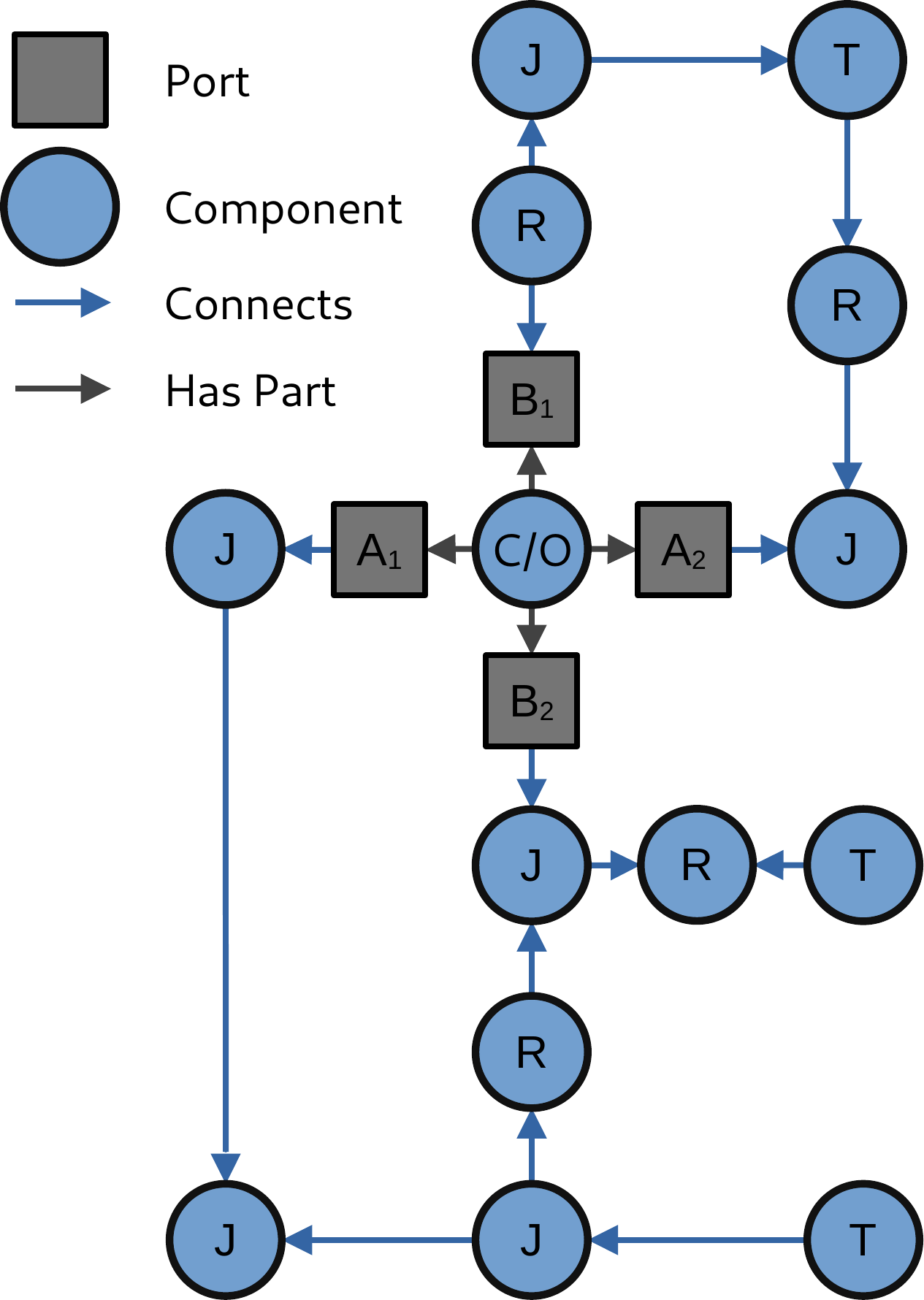}
  \caption{Sample Circuit in Proposed Representation. For visibility reasons, relations to the owing graph resource and the positioning are omitted; port names as well as component types are denoted as abbreviated labels. Note that the direction of component connections is unordered.}
  \label{fig:sample}
\end{figure}

In oder to have a common ontology for describing the circuits, the following terminology is used:

A \textbf{circuit} is described as a graph structure with \textbf{components} as nodes and electrical \textbf{connections} as edges. Components can be either connected directly or via \textbf{ports} which represent the individual terminals of a component (e.g. the anode of a diode). Additionally, \textbf{junctions} and \textbf{crossover} symbols are introduced to support the graphical representation of engineering diagrams as well as their digitization from analogue sources. Junctions are used both to indicate corners in wiring lines and to allow for connecting multiple components (to form hyperedges in graph logic). Crossover symbols indicate the crossing of lines without a electrical connection between them. Components annotation rules add \textbf{functions} to indicate a component's purpose in the circuit. All mentioned terms (including the specific \textbf{compoment classes} and \textbf{function classes}) are related to Wikidata \cite{van2019wikidata} entries in the circuit's RDF representation.

\subsection{Preprocessing Rules}

The preprocessing rules create a normalized electrical view by augmenting the raw RDF structure. Therefore, they allow a simplified and flexible design of the annotation rules:

\subsubsection{Electrical Symmetry}

\begin{figure}
  \centering
  \includegraphics[width=0.2\textwidth]{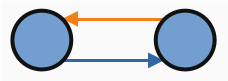}
  \caption{Electrical Symmetry Rule (result in orange).}
  \label{fig:symmetry}
\end{figure}

The RDF triples which express the electrical connections between the components form a directed graph, while there is no physical justification for the level at which the annotation rules are applied. In fact, the position of subject and object resource is considered not well defined in the respective triples (i.e. up to the implementation of the graph generation), resulting in an directed yet unordered relation. In order to implement an undirected graph structure and consequently allow for an abstraction before the annotations rule application, connecting triples of opposite direction are added (see also Figure~\ref{fig:symmetry}):

\begin{lstlisting}
[electSymm: (?a w:connects ?b)
            -> (?b w:connects ?a)]
\end{lstlisting}

\subsubsection{Junction Resolution}

\begin{figure}
  \centering
  \includegraphics[width=0.2\textwidth]{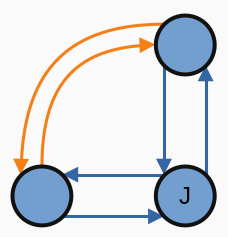}
  \caption{Junction Resolution Rule (results in orange; after prior symmetry rule application).}
  \label{fig:junction}
\end{figure}

As junctions are both optical features and a mean to describe hyperedges (which are tedious to encode in forward chaining rules), they also need to be broken down to direct connections between components. This can be achieved by a simple transitive relation (see Figure~\ref{fig:junction}):

\begin{lstlisting}
[byJ: (?a w:connects ?junction),
      (?junction w:connects ?c),
      (?junction rdf:type w:JUNCTION)
      -> (?a w:connects ?c)]
\end{lstlisting}

\subsubsection{Port Resolution}

\begin{figure}
  \centering
  \includegraphics[width=0.32\textwidth]{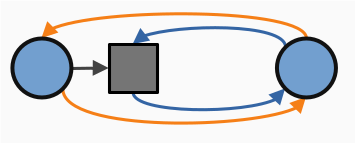}
  \caption{Port Resultion Rule (results in orange; after symmetry rule application).}
  \label{fig:port}
\end{figure}

As information model allows electrical connections between either \textbf{components} or \textbf{ports} or a mixture of them, rules may or may not address them. As circuits of different granularity should be supported and some rules don't make use of the port attributes, it is crucial to add connections so that components are always connected directly. Note that this rule is used in conjunction with the electrical symmetry (see Figure~\ref{fig:port}):

\begin{lstlisting}
[res: (?owner w:has_part ?port),
      (?port rdf:type w:PORT),
      (?a w:connects ?port)
      -> (?a w:connects ?owner)]
\end{lstlisting}

\subsubsection{Crossover Resolution}

\begin{figure}
  \centering
  \includegraphics[width=0.49\textwidth]{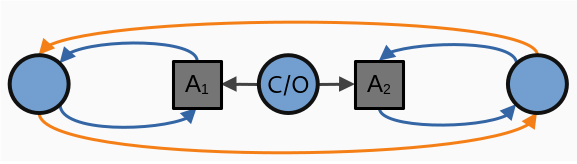}
  \caption{Crossover Resolution Rule (results in orange; after prior symmetry rule application).}
  \label{fig:crossover}
\end{figure}

Crossover symbols are resolved by connecting the connection partners of the opposite crossover symbol ports (the rule below only describes the resolution of one pair of opposite crossover ports while a crossover symbol usually has two pairs, see Figure~\ref{fig:crossover}):

\begin{lstlisting}
[resCro: (?a w:connects ?co_a1),
         (?b w:connects ?co_a2),
         (?co rdf:type w:CROSSOVER),
         (?co_a1 w:name a_1),
         (?co_a2 w:name a_2),
         (?co_a1 rdf:type w:PORT),
         (?co_a2 rdf:type w:PORT),
         (?co w:has_part ?co_a1),
         (?co w:has_part ?co_a2),
         (?junction rdf:type w:JUNCTION)
         -> (?a w:connects ?b)]
\end{lstlisting}

\subsection{Component Annotation Rules}

While the preprocessing rules are considered to be a fixed set, the component annotation rules are an intended to be extensible by domain experts and knowledge workers. The component annotation rules used in this paper are:

\begin{center}
  \begin{tabular}{|c | c|} 
    \hline
    Name & Description \\ 
    \hline
    \hline
    
    Emitter    & A bipolar transistor amplifier\\
    Common     & using a voltage divider biasing\\
    Amplifier  & to keep base bias voltage at a\\
               & constant level.\\
    \hline
    Coupling   & Connects the AC part of a signal\\
    Capacitor  & between two parts of the circuit\\
               & while blocking DC Parts.\\    
    \hline
    Electronic & Component in either open or\\
    Switch     & closed state.\\
    \hline
    Flyback    & A diode that is connected\\
    Diode      & inversely to an energy storage\\
               & component for protecting\\
               & against voltage spikes.\\
    \hline
    Oscillator & Provide a constant stable\\
    Crystal    & frequency and can be used\\
               & as clock in digital circuits\\
    \hline
    PullUp     & Provides a well-defined voltage\\
    Resistor   & level in case of absence of other\\
               & connections (e.g. open switches).\\
    \hline
    Voltage    & provides an intermediate voltage\\
    Divider    & level between the surrounding\\
               & voltage levels.\\
    \hline
    
  \end{tabular}
\end{center}

\subsection{Implementation}

Complete circuits diagrams are loaded from KiCad \cite{kanagachidambaresan2021introduction} schematic files and internally captured as NetworkX \cite{hagberg2020networkx} graphs before being converted to RDF Turtle \cite{turtle} representations. The preprocessing as well as the component annotation itself is performed as forward chaining rules in Apache Jena \cite{siemer2019exploring}, where the component annotation rules are implemented as individual files for extensibility purposes. The source code and the circuit dataset is made publicly available \footnote[1]{https://github.com/DFKI/circuitgraph-insights}.

\section{\uppercase{Evaluation}}
\label{sec:evaluation}

\begin{figure}
  \centering
  \includegraphics[width=0.49\textwidth]{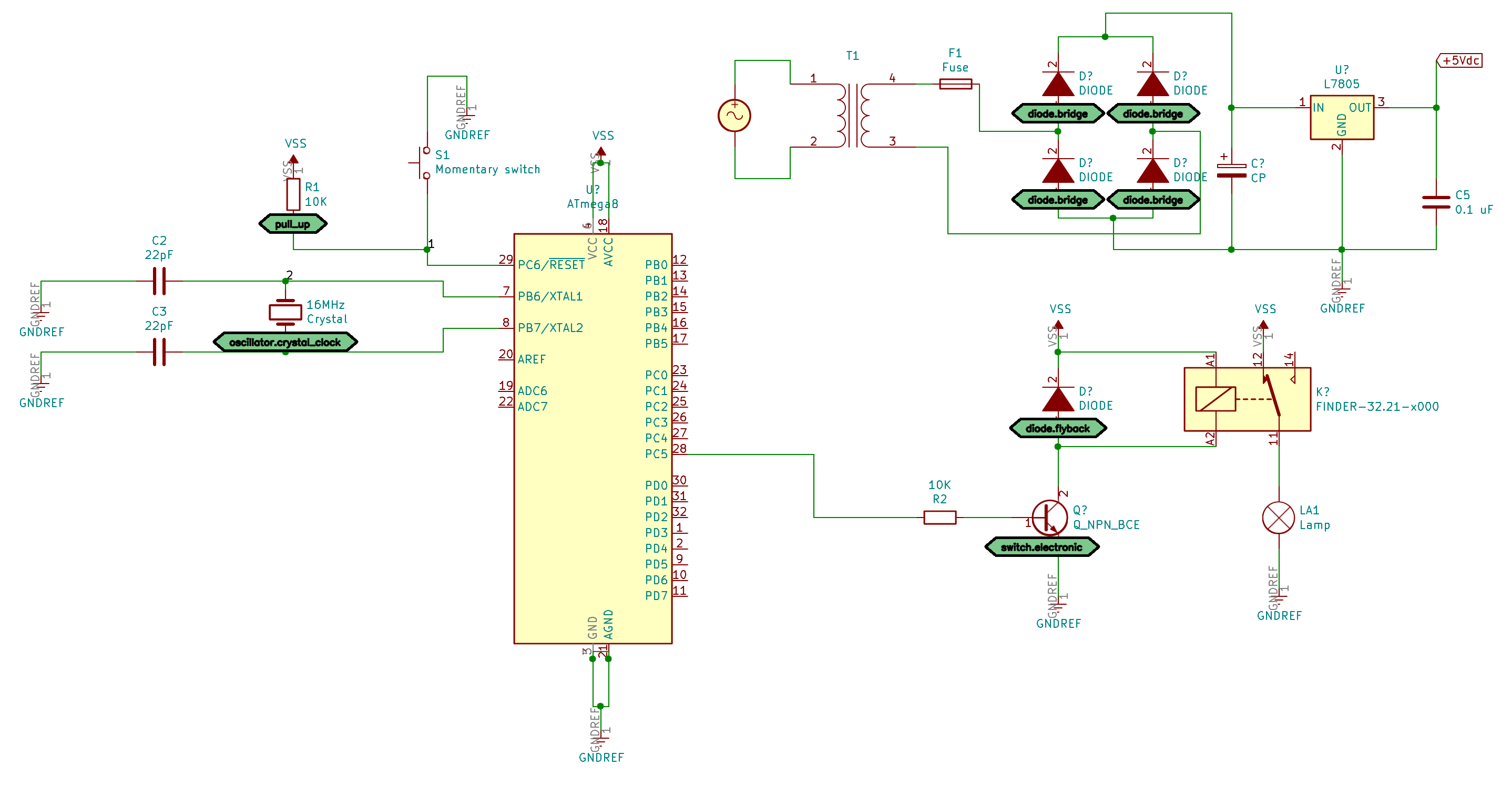}
  \caption{Component Annotations (Green Hexagons) on a Sample Circuit.}
  \label{fig:eval}
\end{figure}

In order to validate the approach, the results are demonstrated on a sample circuit (see Figure~\ref{fig:eval}).

\section{\uppercase{Conclusion}}
\label{sec:conclusion}

An RDF-based system for automatically deriving functional annotations of individual components inside circuits has been described. By incorporating support for an openly available CAE system as well as referencing ressources from the also openly available wikidata knowledge base, it connects the world of circuit modeling with the world of image annotation and the world of circuit understanding.

\section{\uppercase{Outlook}}
\label{sec:outlook}

So far, many of the rules needed to be formulated in multiple, rather specific ways in order to deal with all desired situations. Further preprocessing steps are required to allow for more general formulations. For example, voltage sources as well as vcc and gnd symbols need to be resolved to a uniform representation.

\section*{\uppercase{Acknowledgment}}
\label{sec:Acknowledgement}
This work was funded by the BMBF project SensAI (grant no. 01IW20007).

\bibliographystyle{apalike}
{\small
\bibliography{paper}}

\end{document}